\documentclass[aps,prd,twocolumn,superscriptaddress,groupedaddress,amsmath,amssymb,nofootinbib,preprintnumbers]{revtex4}

\usepackage{graphicx}  % needed for figures
\usepackage[mathlines]{lineno} % cannot get switch to work!
\usepackage[english]{babel} 
\usepackage{xcolor}
\usepackage{ulem}
\usepackage{dcolumn}   % needed for some tables
\usepackage{bm}        % for math
\usepackage{amssymb}   % for math
\usepackage{amsbsy}   % for math
\usepackage{hyperref}
\usepackage{nicefrac}
\usepackage{todonotes}
\usepackage{comment}

\hypersetup{
    colorlinks=true,
    linkcolor=cyan,
    filecolor=magenta,      
    urlcolor=cyan,
    citecolor=violet,
}

\makeatletter
\g@addto@macro\bfseries{\boldmath}
\makeatother
\begin{document}

\title{Features of the energy spectrum of cosmic rays above 
$2.5{\times} 10^{18}$~eV using the\\ Pierre Auger Observatory}
%\input author_list.tex       % D0 authors (remove the first 3 lines
                             % of this file prior to submission, they
                             % contain a time stamp for the authorlist)
                             % (includes institutions and visitors)
%\date{\today}

% created on 2020-05-05

\author{A.~Aab}
\affiliation{IMAPP, Radboud University Nijmegen, Nijmegen, The Netherlands}

\author{P.~Abreu}
\affiliation{Laborat\'orio de Instrumenta\c{c}\~ao e F\'\i{}sica Experimental de Part\'\i{}culas -- LIP and Instituto Superior T\'ecnico -- IST, Universidade de Lisboa -- UL, Lisboa, Portugal}

\author{M.~Aglietta}
\affiliation{Osservatorio Astrofisico di Torino (INAF), Torino, Italy}
\affiliation{INFN, Sezione di Torino, Torino, Italy}

\author{J.M.~Albury}
\affiliation{University of Adelaide, Adelaide, S.A., Australia}

\author{I.~Allekotte}
\affiliation{Centro At\'omico Bariloche and Instituto Balseiro (CNEA-UNCuyo-CONICET), San Carlos de Bariloche, Argentina}

\author{A.~Almela}
\affiliation{Instituto de Tecnolog\'\i{}as en Detecci\'on y Astropart\'\i{}culas (CNEA, CONICET, UNSAM), Buenos Aires, Argentina}
\affiliation{Universidad Tecnol\'ogica Nacional -- Facultad Regional Buenos Aires, Buenos Aires, Argentina}

\author{J.~Alvarez Castillo}
\affiliation{Universidad Nacional Aut\'onoma de M\'exico, M\'exico, D.F., M\'exico}

\author{J.~Alvarez-Mu\~niz}
\affiliation{Instituto Galego de F\'\i{}sica de Altas Enerx\'\i{}as (IGFAE), Universidade de Santiago de Compostela, Santiago de Compostela, Spain}

\author{R.~Alves Batista}
\affiliation{IMAPP, Radboud University Nijmegen, Nijmegen, The Netherlands}

\author{G.A.~Anastasi}
\affiliation{Universit\`a Torino, Dipartimento di Fisica, Torino, Italy}
\affiliation{INFN, Sezione di Torino, Torino, Italy}

% ...

\collaboration{The Pierre Auger Collaboration}
\email{auger_spokespersons@fnal.gov}
\homepage{http://www.auger.org}
\noaffiliation

\begin{abstract}

We report a measurement of the energy spectrum of cosmic rays above $2.5{\times} 10^{18}$~eV based on $215,030$ events.  New results are presented: at about  $1.3{\times} 10^{19}$~eV, the spectral index changes from $2.51 \pm 0.03 \textrm{ (stat.)}  \pm 0.05 \textrm{ (sys.)}$ to $3.05 \pm 0.05 \textrm{ (stat.)}\pm 0.10\textrm{ (sys.)}$, evolving to $5.1\pm0.3\textrm{ (stat.)} \pm 0.1\textrm{ (sys.)}$ beyond $5{\times} 10^{19}$~eV, while no significant dependence of spectral features on the declination is seen in the accessible range. These features of the spectrum can be reproduced in models with energy-dependent mass composition. The energy density in cosmic rays above $5{\times} 10^{18}$~eV  is  $(5.66 \pm 0.03 \textrm{ (stat.)} \pm 1.40 \textrm{ (sys.)} ) {\times} 10^{53}~$erg~Mpc$^{-3}$.

\end{abstract}

\maketitle

%\textbf{General guideline, from PRL site}
%\begin{itemize}
%\item 3750 words
%\item Title: avoid hyperboles assessing the quality of the work (``precise'', ``important'', or ``accurate'')
%\item Abstract: 600 characters at most
%\item No sections -- but ``virtual sections'' with ``titles'' in italics 
%\item Joint submission with the long paper, so the paper and the Letter should cite each other.
%\item Figures should have a width of a 8.6 cm 
%\end{itemize}

%\textbf{Synopsis}\\

%1.    
Although cosmic rays having energies above $10^{19}~$eV were first detected nearly 60 years ago~\cite{LSR1961,Linsley1963} and are being investigated by the two largest-ever built detectors, the Pierre Auger Observatory~\cite{ThePierreAuger:2015rma} and the Telescope Array~\cite{TA}, the question of their origin remains unanswered.  Only recently has the belief that such particles are of extragalactic origin been demonstrated experimentally with the discovery of significant directional anisotropies above $8{\times} 10^{18}~$eV~\cite{Auger2017Science}.  These data are well-described by a dipole pattern, the amplitude of which increases from $\simeq 6\%$ to $\simeq 10\%$ as the energy rises to $\simeq 4{\times} 10^{19}~$eV~\cite{Auger2018ApJ}. 

%2
An important observable for an understanding of ultra-high energy cosmic rays (UHECRs) is the energy spectrum.  We report a measurement above $2.5 {\times} 10^{18}~$eV based on 215,030 events, over 10 times that used in~\cite{Auger2008PRL}. Over 16,000 events have energies beyond $10^{19}~$eV.  This spectral determination is unique in making no assumptions about the mass composition or the hadronic physics. Full details are reported in~\cite{Auger2019PRD}.

%3.    
UHECRs can only be studied through the detection of the showers of particles (extensive air-showers) they create in the atmosphere.  A calorimetric estimate of the energy carried by the primary particle is possible using telescopes to collect the fluorescence light emitted by atmospheric nitrogen excited by the shower.  
The on-time of this technique is limited to nights with low-background light while, by contrast, an array of particle detectors deployed on the ground can be operated with a duty-cycle close to 100\%.  The traditional method of assessing the energy of the primary cosmic-ray from observations made with the particle detectors requires assumptions about its mass and the hadronic processes that control the cascade development.  This is clearly unsatisfactory as the mass is unknown and the centre-of-mass energy reached at the LHC corresponds only to that of a proton of $\simeq 10^{17}~$eV colliding with a nitrogen nucleus.  Also, details of pion interactions, which play a key role in shower development, are lacking.  The presence of unknown processes could also lead to hidden systematic uncertainties.

%4.    
To circumvent these limitations, the energies are obtained by making use of a subset of events detected simultaneously by the fluorescence detector (FD) and the particle surface detectors (SD).  This ``hybrid'' approach allows a calorimetric estimate of the energy for events recorded during periods when the FD cannot be operated.  A spectrum can thus be derived that is free from assumptions about primary mass or hadronic physics. 

%5. 
The Pierre Auger Observatory is such a hybrid system~\cite{ThePierreAuger:2015rma}. It is sited near the city of Malarg\"ue, Argentina, at latitude 35.2$^{\circ}$~S with a mean atmospheric overburden of 875~g~cm$^{-2}$~\cite{ThePierreAuger:2015rma}. The SD comprises 1600 water-Cherenkov detectors deployed on a 1500~m triangular grid, covering about 3000~km$^{2}$. The array is overlooked from four stations, each containing six telescopes used to detect the emitted fluorescence light. Comprehensive atmospheric monitoring, particularly of the aerosol content and the cloud cover, is undertaken~\cite{ThePierreAuger:2015rma,Abreu:2012zg}.

%6.    
The SD samples the shower particles that reach the ground. Signals in the individual detectors are quantified in terms of their response to a muon travelling vertically and centrally through it (a vertical equivalent muon or VEM). The signals are used to determine the impact point of the shower axis, the arrival direction and the shower size. For the latter, the signal at 1000~m from the shower axis, $S(1000)$, is used. For the grid spacing of 1500~m, this is the distance that minimizes the uncertainty arising in $S(1000)$ from the imperfect knowledge of the functional form describing the fall-off of signal with distance from the shower axis in individual events~\cite{NKW2007}. 

%7.    
Showers detected by the SD arrive from a range of zenith angles, and they are attenuated according to how much atmosphere is traversed. Accordingly, for each event, $S(1000)$ is adjusted to a reference value, $S_{38}$, the magnitude that it would have had, had the cosmic ray arrived at the median zenith angle of $38^\circ$. The long-established procedure for making this correction, the Constant Intensity Method~\cite{Hersil1961}, relies on the quasi-isotropy of cosmic rays in zenith angle given the small anisotropy contrasts in celestial coordinates~\cite{Auger2019PRD}.  The large number of events has made it possible to refine the original approach and quantify the change in shower absorption as a function of energy. Such an evolution is anticipated as, at a given zenith angle, the ratio of the muon to electromagnetic components falls as the energy increases, even for an energy-independent composition. 

%8.    
For showers detected by the FD, it is possible to measure the deposition of energy lost to ionisation of the atmosphere using a fit to a modified Gaisser-Hillas profile~\cite{AugerLR2019}. The integration of the profile yields a calorimetric measure of this loss. The energy of the primary particle, $E_{\mathrm{FD}}$, is then obtained by the addition of an energy-dependent correction of less than 14\%, driven by data~\cite{InvisibleEnergy2019}, to allow for the ``invisible energy'', carried into the ground by muons and neutrinos. The resolution in $E_{\mathrm{FD}}$ is well-described by $\sigma_{\textrm{FD}}(E)/E\simeq 7.4\%$ over the whole energy range~\cite{DawsonICRC2019}.

%9.    
Hybrid events are thus used to develop a calibration curve such that every estimate of $S_{38}$ can be assigned a valuation of $E_{\mathrm{FD}}$.  Here 3,338 hybrid events surviving rigorous quality cuts~\cite{Auger2019PRD} are used to obtain a relationship between $S_{38}$ and $E_{\mathrm{FD}}$ of the form $E_\textrm{FD} = A~{S_{38}}^{B}$, where $A=(1.86 \pm 0.03){\times} 10^{17}$~eV and $B=1.031 \pm 0.004$.  No zenithal dependence of $A$ or $B$ has been found, further validating the use of the Constant Intensity Method~\cite{Auger2019PRD}.  Such a simple dependence is sufficient to describe the data in full detail. The energies of the hybrid events range from $2.5 {\times} 10^{18}$~eV to $8{\times} 10^{19}$~eV. The most energetic event, detected at all fluorescence stations, has an energy $E_\textrm{FD} = (8.5 \pm 0.4) {\times} 10^{19}$~eV, derived from a weighted average of the four independent estimates of the calorimetric energy. For this event $S_{38} = 354$~VEM so that the energy deduced from the calibration curve is $E_{\mathrm{SD}}\equiv A{S_{38}}^B = (7.9 \pm 0.6) {\times} 10^{19}$~eV. The systematic uncertainty in the energy assignment is about 14\% over the whole energy range~\cite{Verzi2013}. This  benefits from the high-precision AIRFLY measurement of the fluorescence yield~\cite{FY-Airfly_AbsYield} and from an accurate data-driven estimation of the invisible energy~\cite{InvisibleEnergy2019}. Other contributions to the uncertainty are related to the estimation of the $A$ and $B$ parameters, the characterization of the atmosphere, the reconstruction of the longitudinal profile and the FD calibration, which provides the largest contribution. 

%10
To derive the energy spectrum, we use events recorded by the SD with the largest-signal station not located on the boundary of the array, with zenith angle $\theta<60^\circ$ and energy $\geq2.5{\times} 10^{18}$~eV. These selection criteria not only ensure adequate sampling of the shower but also allow the evaluation of the aperture of the SD in a purely geometrical manner in the regime where the array trigger is fully efficient and independent of the mass or energy of the primary particle~\cite{Abraham:2010zz}. The resulting SD data set consists of 215,030 events recorded between 1 January 2004 and 31 August 2018, from an exposure, $\mathcal{E}$, of $(60{,}400 \pm 1{,}810)$ \,km$^2$\,sr\,yr. The determination of $\mathcal{E}$, dependent only on the acceptance angle, the surface area and the live-time of the array, is discussed in detail in~\cite{Abraham:2010zz}.

%11
The procedure for extracting the spectrum from the observations, fully discussed in~\cite{Auger2019PRD}, is summarised here.  

%12
The energy spectrum, typically a power law ($\propto E^{-\gamma}$) with spectral index $\gamma$ in a given energy interval, is estimated as $J_i=c_i N_i/\left(\mathcal{E}\Delta E_i\right)$, with $N_i$ the number of observed events in differential bins of width $\Delta \log_{10}E_i=0.1$ and $c_i$ the correction factors required to eliminate the biases caused by the finite energy resolution. The size of the bins is such that it corresponds approximately to the energy resolution in the lowest energy bin, which starts at $2.5{\times} 10^{18}$ eV.

%13
The correction factors are needed because, as the spectrum is steep, the finite resolution causes migration between bins, particularly from lower to higher energies, artificially enhancing the flux. At the lowest energies, the correction depends also on the behaviour of the detection efficiency in the energy region where the array is not fully efficient as well as on the bias in the energy due to trigger-selection effects.%the higher efficiency to detect showers induced by heavier primary nuclei~\cite{Abraham:2010zz}.

%14
A forward-folding approach is used to determine the correction factors. It consists of finding the model of the energy spectrum folded  for detector effects that best describes the data, and then using this model to calculate the values of $c_i$. The SD efficiency can be estimated from the fraction of hybrid events that also satisfy the SD trigger conditions, because above $10^{18}$~eV, the hybrid trigger efficiency is 100\% independent of primary mass~\cite{ExpoHybrid2011}. The energy resolution of $E_{\mathrm{SD}}$, and the bias in its estimate, are found from a study of the distributions of $E_{\mathrm{SD}}/E_{\mathrm{FD}}$. The resolution improves from $\approx 20\%$ at $2{\times} 10^{18}$~eV to $\approx 7\%$ at $2{\times} 10^{19}$~eV and is constant thereafter. 
The bias is zero above $2.5{\times} 10^{18}$~eV and increases smoothly going to lower energies and larger zenith angles: at $10^{18}$~eV it is $\approx 10\%$ at $0^\circ$ and $\approx 30\%$ at $60^\circ$.

%15
Thanks to the hybrid measurements, the correction factors are estimated avoiding any reliance on model and primary mass assumptions. The factors  are maximal at the lowest energies, $\approx 8\%$, and less than 5\% at the highest energies available. Further details are given in~\cite{Auger2019PRD}.

%\begin{widetext} 
%   	\begin{center}
  %%%%%%%%%%%%%%%%%%%%%%
   %%%%%%%%%%%%%%%%%%%%%%
   \begin{figure}[t]
   	\centering
   	\includegraphics[width=8.6cm]{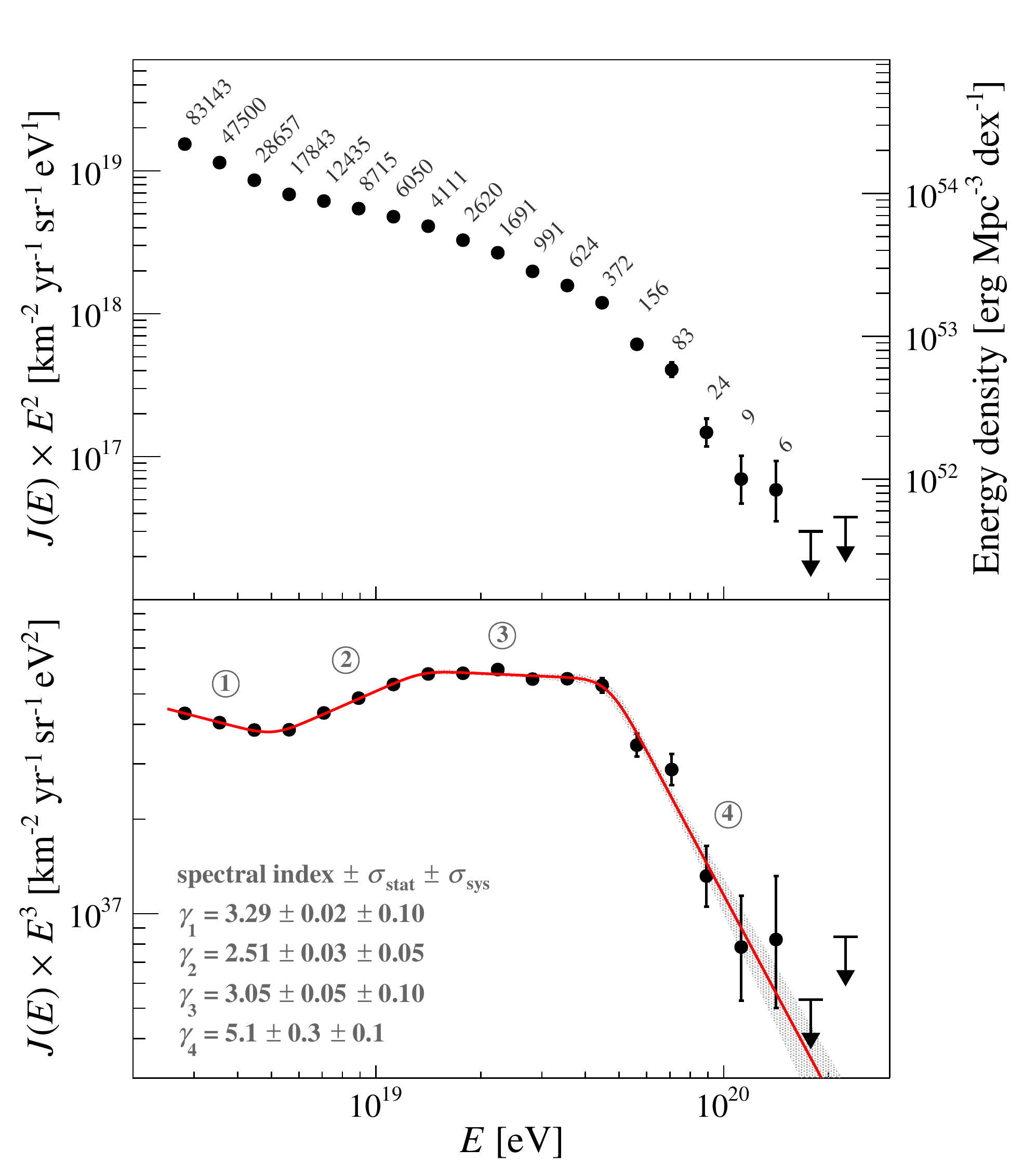}
   	\caption{\small{Top: energy spectrum scaled by $E^{2}$ with the number of detected events in each energy bin. In this representation the data provide an estimation of the differential energy density per decade. Bottom: energy spectrum scaled by $E^{3}$ fitted with a sequence of four power laws (red line). 
           The numbers ($i = 1, \ldots, 4$) enclosed in the circles identify the energy intervals where the spectrum is described by a power law with spectral index $\gamma_i$.
   	   The shaded band indicates the statistical uncertainty of the fit.
   	   Upper limits are at the $90\%$ confidence level.}} 
   	\label{fig:spectrum}
   \end{figure}
   %%%%%%%%%%%%%%%%%%%%%%
   %%%%%%%%%%%%%%%%%%%%%%
%   	\end{center}
%\end{widetext}

%16
The model of the energy spectrum that we used for over a decade is a series of two power laws followed by a slow suppression. With the current exposure, this model turns out to describe the data poorly, as the reduced deviance is found to be $35.6/15$~\cite{Auger2019PRD}. Consequently, we adopt a more complex function with a sequence of four power laws with smooth transitions~\cite{Lipari2018}, 
\begin{equation}
J(E)=J_0\left(\frac{E}{10^{18.5}~\textrm{eV}}\right)^{-\gamma_1}\prod_{i=1}^3\left[1+\left(\frac{E}{E_{ij}}\right)^{\frac{1}{\omega_{ij}}}\right]^{(\gamma_i-\gamma_j)\omega_{ij}}\hspace{-0.15cm},\nonumber
\label{eqn:J}
\end{equation}
with $j=i+1$ and $\omega_{ij}=0.05$. The $\omega_{ij}$ factors control the widths of the energy intervals over which the slope transitions occur~\cite{Auger2019PRD}. This model describes the data with a reduced deviance $17.0/12$, which allows us to disfavor the previous parameterization with $3.9\sigma$ confidence~\cite{Auger2019PRD}. The resulting differential energy spectrum and the fitted function are shown in Fig.~\ref{fig:spectrum}.
The normalization is $J_0=(1.315 \pm 0.004 \pm 0.400) {\times} 10^{-18}$~km$^{-2}$~sr$^{-1}$~yr$^{-1}$~eV$^{-1}$. 
The ankle is described by a rollover at $E_{12} = (5.0\pm0.1\pm0.8) {\times} 10^{18}$~eV, marking a hardening of the spectrum from $\gamma_1 = 3.29\pm0.02\pm0.10$ to $\gamma_2 = 2.51\pm0.03\pm0.05$. At $E_{23} = (13\pm1\pm2) {\times} 10^{18}$ eV, the spectrum softens from $\gamma_2$ to $\gamma_3 = 3.05\pm0.05\pm0.10$. 
Finally, the spectrum softens further above a suppression energy of $E_{34} = (46\pm3\pm6) {\times} 10^{18}$ eV with $\gamma_4 = 5.1\pm0.3\pm0.1$, confirming with higher precision previous reports of the strong attenuation of the flux at the highest energies~\cite{HIRES,Auger2008PRL,TA2012}. The feature at $E_{23}$, calling for a 2-step suppression, is a new observation. 
For all parameters and observables presented in the text, the first error is statistical and the second systematic. 

From the measured  energy spectrum one can infer the differential energy density per dex\footnote{dex indicates decade in $\log_{10}E$, following the convention of~\cite{dex1951}.}, obtained as $\ln\left(10\right) \left(4 \pi /c \right) E^2  J(E)$. It provides a measurement of the energy density of the local Universe attributable to cosmic rays. Above the ankle, a range in which UHECRs are of extragalactic origin~\cite{Auger2017Science}, the integration over energy results in $(5.66 \pm 0.03 \pm 1.40) {\times} 10^{53}~$erg~Mpc$^{-3}$.
This translates into constraints on the luminosity of the sources, as discussed below.

%17
A detailed examination of the systematic uncertainties of the energy spectrum is reported in~\cite{Auger2019PRD}.  The uncertainty in the flux amounts to 30 -- 40\% near $2.5 {\times} 10^{18}$~eV, 25\% at $10^{19}$~eV and 60\% at the highest energies. The uncertainties include contributions from the absolute energy scale (the largest), the exposure, the unfolding procedure and the $S(1000)$ reconstruction. No indication of further systematic uncertainties has been found from a comparison of the spectra calculated over different time periods, seasons and ranges of zenith angle.

\begin{table}[h]
\caption{Spectral parameters in three different declination ranges. The energies $E_{12}$, $E_{23}$ and $E_{34}$ are given in units of $10^{18}~$eV
and the normalization parameter $J_0$ in units of $10^{18}$~km$^{-2}$sr$^{-1}$yr$^{-1}$eV$^{-1}$. Uncertainties are statistical.}
\label{tab:spectralfeatures_dec}
\begin{ruledtabular}
\begin{tabular}{l c c c}
 &  $[-90.0^\circ,-42.5^\circ]$ & $[-42.5^\circ, -17.3^\circ]$ & $[-17.3^\circ, +24.8^\circ]$ \\
\colrule
$J_0$ & $1.329 \pm 0.007$ & $1.306 \pm  0.007$ & $1.312 \pm 0.006 $ \\  
$\gamma_1$              & $3.26\pm0.03$  & $3.31\pm0.03$  & $3.30\pm0.03$ \\
$\gamma_2$              & $2.53\pm0.04$  & $2.54\pm0.04$  & $2.44\pm0.05$ \\
$\gamma_3$              & $3.1\pm0.1$    & $3.0\pm0.1$    & $3.0\pm0.1$ \\
$\gamma_4$              & $5.2\pm0.4$    & $4.4\pm0.3$    & $5.7\pm0.6$  \\
$E_{12}~$               & $5.1\pm0.2$    & $4.9\pm0.2$    & $5.2\pm0.2$ \\
$E_{23}~$               & $14\pm2$       & $14\pm2$       & $12\pm1$ \\
$E_{34}~$               & $47\pm4$       & $37\pm4$       & $51\pm4$
\end{tabular}
\end{ruledtabular}
\end{table}

%18
The wide declination range covered, from $\delta=-90^\circ$ to $\delta=+24.8^\circ$, allows a search for dependencies of energy spectra on declination. For this, we have divided the sky into three declination bands of equal exposure.  In each band, the estimation of the spectrum is made as for the whole field of view, but using unfolding-correction factors relevant to the band in question. We report in Table~\ref{tab:spectralfeatures_dec} the parameters characterizing the spectral features for each declination range. They are seen to be in statistical agreement. There is thus no obvious dependence with declination over the energy range covered. 
A trend for the intensity to be slightly higher in the Southern Hemisphere is observed~\cite{Auger2019PRD}, consistent with the anisotropy observations~\cite{Auger2018ApJ}.  We therefore claim a second new result, namely that the energy spectrum does not vary as a function of declination in the range accessible at the Auger Observatory other than in the mild excess from the Southern Hemisphere expected in line with the known energy-dependent anisotropies above $8{\times} 10^{18}$~eV. A comparison of the spectrum with that of Telescope
Array measured in the Northern hemisphere is discussed in~\cite{Auger2019PRD} and references therein.

%19
\textit{Astrophysical implications of the features of the energy spectrum.} We now examine the validity of models proposed to explain features of UHECRs using the new information given here and the data on mass composition and arrival directions recently reported~\cite{Auger2017Science,Auger2018ApJ,Auger2014aPRD,Auger2014bPRD,YushkovICRC2019,Auger2016PLB,Auger2018ApJL,Caccianiga2019}. If UHECRs are produced throughout the Universe, to reach Earth they must cross the background photon fields permeating the extragalactic space. In particular, the cosmic microwave background photons induce pion-production with protons colliding at around $5{\times} 10^{19}$~eV and photo-disintegration of heavier nuclei at a roughly similar threshold, leading to the expectation of a spectral steepening (the Greisen-Zatsepin-Kuz’min (GZK) effect~\cite{GZK}). Depending on the energy and chemical composition of the UHECRs, higher-energy background photons, such as infrared light, may also be responsible of interactions producing the flux steepening. 

%20
A popular framework has been that what is observed comes from universal sources, uniformly distributed, that accelerate only protons.  As a consequence the ankle is then explicable by energy losses of protons through pair production across greater distances~\cite{Hillas1967,BGG2005,BGG2006} so that the ankle region would be proton-dominated.  However, recent results~\cite{Auger2016PLB} strongly contradict this expectation: in the ankle region, $(3-5){\times} 10^{18}$~eV, a pure proton composition, or one of only protons and helium, is excluded at the $6.4~\sigma$ level. A second consequence~\cite{BG1988} concerns the energy, $E_{1/2}$, at which the integral intensity falls by a factor two with respect to a power-law extrapolation from lower energies. The prediction in this framework is that $E_{1/2}=5.3{\times} 10^{19}$~eV though this number may change with fluctuations of source luminosities and densities that shape the GZK feature~\cite{BGG2005,BBO2001}, and with the maximum achievable energy in the sources. 
The value found here, $(2.2 \pm 0.1  \pm 0.3){\times} 10^{19}$~eV, 
is at variance with the prediction because of the new feature of the spectrum at~$\approx 10^{19}$~eV, which is absent in the popular paradigm that is thus disfavored.

%21
Relaxing the universality of the source spectra, the steepening at $\approx 10^{19}$~eV could stem from the distinctive spectrum of a local source that emits protons and contributes significantly to the total intensity. At these energies, diffusive propagation of protons from a nearby source is excluded by limits set on extragalactic magnetic fields from rotation measures~\cite{PTU2016}. Approximately, protons would thus arrive to the Galaxy as a uniform, parallel beam that may subsequently be focused or defocused while propagating in the Galactic magnetic field. As seen from the Earth, the image of the source is expected to be shifted and broadened, with the effect growing with decreasing energy. Also, multiple broad images may be produced if uncorrelated regions of the magnetic field are experienced by the particles~\cite{Harari2002,Farrar2014,Keivani2014}. Such a scenario would thus imply the observation of an anisotropy at intermediate angular scales, the size of which depends on the model of turbulence for the magnetic field~\cite{Farrar2017}. Spectral differences would also consequently appear in some parts of the sky. The softening at $\approx 10^{19}$~eV, in particular, would not be expected in every declination range. The absence of such dependence accordingly disfavors the interpretation that the steepening is due to a source in the local Universe emitting protons. Furthermore, the interplay between the luminosity of a given source and its flux attenuation with distance requires fine-tuning to make viable a scenario in which several sources would emit protons with a distinctive spectrum while at the same time no directional effect would be imprinted upon the observed intensity.

   %%%%%%%%%%%%%%%%%%%%%%
   %%%%%%%%%%%%%%%%%%%%%%
   \begin{figure}[h]
   	\centering
   	\includegraphics[width=8.6cm]{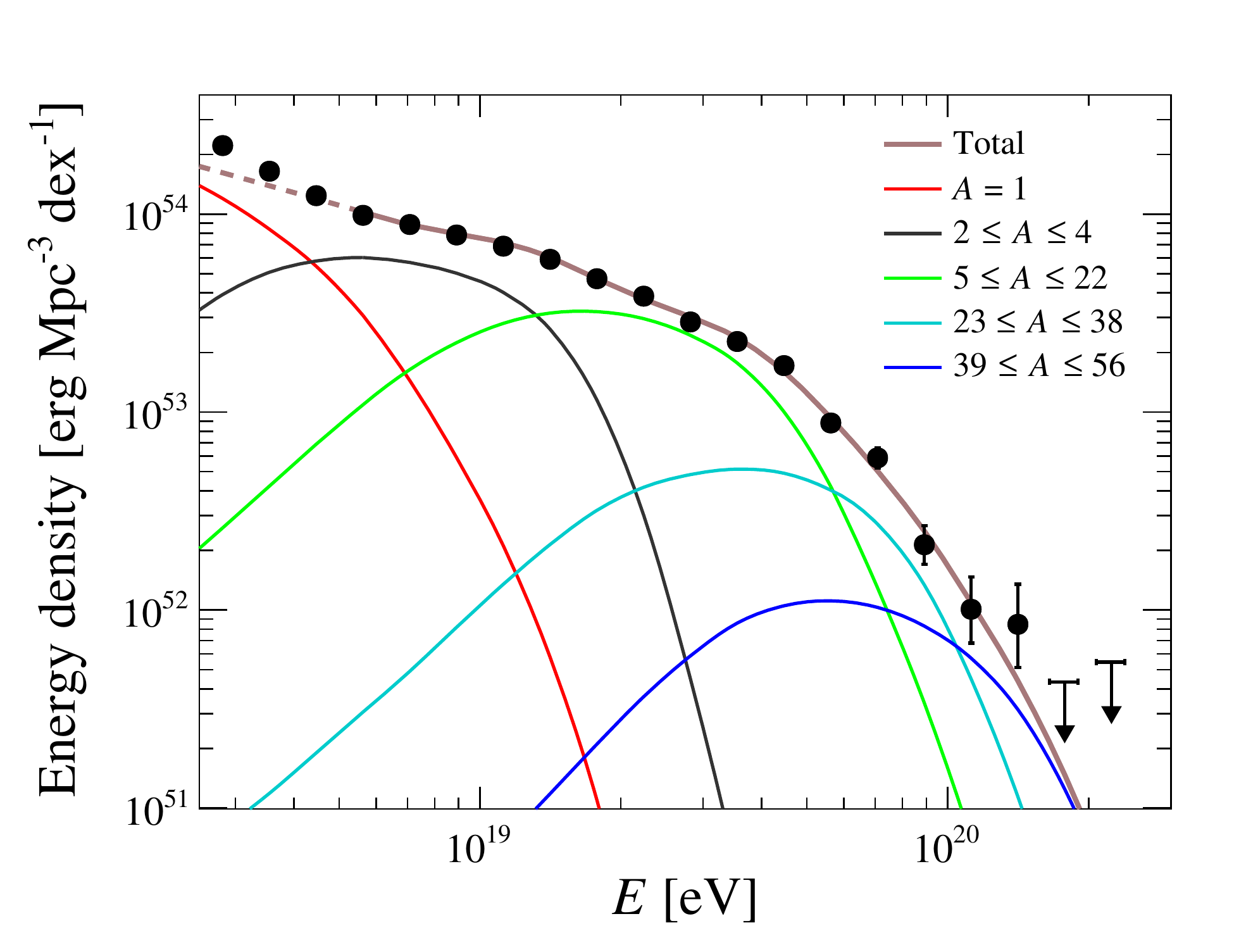}
   	\caption{\small{
   	%Reproduction of the 
   	Energy density obtained with the best fit parameters of the benchmark scenario used for illustration, as described in the text. The dashed curve shows the energy range that is not used in the fit and where an additional component is needed for describing the spectrum.}
   	}
   	\label{fig:spectrum_model}
   \end{figure}
   %%%%%%%%%%%%%%%%%%%%%%
   %%%%%%%%%%%%%%%%%%%%%%

%22
By contrast, our results fit a scenario in which several nuclear components contribute to the total intensity and in which the electromagnetic fields permeate source environments where nuclei are accelerated to a maximum energy proportional to their charge ($Z$). This scenario, e.g.,~\cite{Allard:2007gx,ABB2014,TAH2015,Auger2017JCAP}, provides a natural framework to explain the tendency towards heavier masses with increasing energy as inferred from recent works~\cite{Auger2014aPRD,Auger2014bPRD,YushkovICRC2019}. To illustrate the main physics aspects without distraction by the many details a full model scenario would require, we consider here, as in~\cite{Auger2017JCAP}, several nuclear components injected at the sources with a power-law spectrum and with the maximal energy of the sources modeled with an exponential cut-off. For simplicity, the sources are assumed to be stationary and uniform in a co-moving volume. We show in Fig.~\ref{fig:spectrum_model} the best reproduction of the data by simultaneously fitting the energy spectrum above $5{\times}10^{18}~$eV and the distribution of the depths of the shower maximum ($X_{\rm max}$), which is mass-sensitive
(using EPOS LHC~\cite{TP2015} as model of hadronic interactions in their interpretation).
The abundance of nuclear elements at the sources is dominated by intermediate-mass nuclei accelerated to $\approx 5~Z {\times} 10^{18}~$eV and escaping from the source environments with a very hard spectrum. In this scenario, the steepening observed above $\approx 5{\times} 10^{19}$~eV results from the combination of the maximum energy of acceleration of the heaviest nuclei at the sources and the GZK effect. The steepening at $\approx 10^{19}$~eV reflects the interplay between the flux contributions of the helium and carbon-nitrogen-oxygen components injected at the source with their distinct cut-off energies, shaped by photodisintegration during the propagation. We note that the ratio 
$E_{34}/E_{23}$ is $3.4 \pm 0.3$,  
matching the mass/charge ratio of CNO to He, as expected from the benchmark scenario shown in Fig.~\ref{fig:spectrum_model}. 

Some cautionary comments on the illustrative model considered here are in order. The presence of a sub-dominant light component at the highest energies is not excluded by our data, see e.g.~\cite{muzio2019}. Also, viable source scenarios can be found without resorting to a mixed composition with a rigidity-dependent maximum energy if, for instance, predominately heavy (Si to Fe) nuclei are accelerated and  photo-disintegrate in the source environment~\cite{UFA2015} or en route to Earth~\cite{Hooper:2009fd, Taylor:2011ta}. Scenarios with a predominantly light composition~\cite{BGG2006,Aloisio:2017kpj} can fit our $X_\textrm{max}$ data as well as those of Telescope Array~\cite{Hanlon:2019onl} in the ankle energy range, but these scenarios are at odds with measurements of the correlation of particle densities at ground and $X_\textrm{max}$~\cite{Auger2016PLB}. At ultra-high energies, a significant re-adjustment of current hadronic interaction models would be required~\cite{Farrar:2019cid} to fit our data with a p/He-dominated model while the data of Telescope Array, because of limited statistical power above $10^{19}$~eV, cannot yet be used to draw reliable conclusions about composition in this energy range~\cite{Hanlon:2018hlv}.

Interactions of the accelerated nuclei in the environment of the sources may give rise to copious fluxes of nucleons below the ankle energy, produced through photo-disintegration. 
Neutrons escaping from the magnetic confinement regions may then explain the observed flux of protons deduced from $X_{\rm max}$ measurements~\cite{Auger2014bPRD,YushkovICRC2019} in this energy range, due to neutron decay during propagation~\cite{UFA2015,globus2015a,biehl2018,zhang2018,supanitsky2018,boncioli2019}. To make up the all-particle spectrum and to fit the composition data below $\approx 5{\times} 10^{18}$~eV, an additional component is further  required (see e.g.~\cite{Haungs2015,DelignyCompteRendu,Kachelriess2019} for discussions). This could be the high-energy tail from the sources emitting the bulk of Galactic cosmic rays of lower energies or, as in the ``B-component scenario''~\cite{Hillas2005}, further explored in~\cite{gaisser2013,Thoudam:2016syr}, this additional high-energy component is produced by different sources in the Galaxy.

%23
Finally, within this scenario, our data constrain the luminosity density that continuously emitting sources must inject into extragalactic space in UHECRs to supply the observed energy density. This amounts to $\approx 6 {\times} 10^{44}$~erg~Mpc$^{-3}$~yr$^{-1}$ above $5 {\times} 10^{18}~$eV at a redshift of zero, in line with the value of $2{\times} 10^{44}$~erg~Mpc$^{-3}$~yr$^{-1}$ that can be inferred dividing the measured energy density by the typical cosmic-ray energy loss time 
$O (1 {\rm ~Gpc/c})$   ($3.3 {\rm ~Gyr}$) \cite{Allard:2011aa}. 
%$\mathcal{O} (1 {\rm ~Gpc/c})$   ($3.3 {\rm ~Gyr}$) \cite{Allard:2011aa}. 
Classes of extragalactic sources that match such rates in the gamma-ray band include active galactic nuclei and starburst galaxies~\cite{Dermer2010}. The flux pattern from these objects also provides an indication of anisotropy in UHECR arrival directions~\cite{Auger2018ApJL,Caccianiga2019}.

% created on 2020-05-05

\section*{Acknowledgments}

\begin{sloppypar}
The successful installation, commissioning, and operation of the Pierre
Auger Observatory would not have been possible without the strong
commitment and effort from the technical and administrative staff in
Malarg\"ue. We are very grateful to the following agencies and
organizations for financial support:
\end{sloppypar}

\begin{sloppypar}
Argentina -- Comisi\'on Nacional de Energ\'\i{}a At\'omica; Agencia Nacional de
Promoci\'on Cient\'\i{}fica y Tecnol\'ogica (ANPCyT); Consejo Nacional de
Investigaciones Cient\'\i{}ficas y T\'ecnicas (CONICET); Gobierno de la
Provincia de Mendoza; Municipalidad de Malarg\"ue; NDM Holdings and Valle
Las Le\~nas; in gratitude for their continuing cooperation over land
access; Australia -- the Australian Research Council; Brazil -- Conselho
Nacional de Desenvolvimento Cient\'\i{}fico e Tecnol\'ogico (CNPq);
Financiadora de Estudos e Projetos (FINEP); Funda\c{c}\~ao de Amparo \`a
Pesquisa do Estado de Rio de Janeiro (FAPERJ); S\~ao Paulo Research
Foundation (FAPESP) Grants No.~2019/10151-2, No.~2010/07359-6 and
No.~1999/05404-3; Minist\'erio da Ci\^encia, Tecnologia, Inova\c{c}\~oes e
Comunica\c{c}\~oes (MCTIC); Czech Republic -- Grant No.~MSMT CR LTT18004,
LM2015038, LM2018102, CZ.02.1.01/0.0/0.0/16{\textunderscore}013/0001402,
CZ.02.1.01/0.0/0.0/18{\textunderscore}046/0016010 and
CZ.02.1.01/0.0/0.0/17{\textunderscore}049/0008422; France -- Centre de Calcul
IN2P3/CNRS; Centre National de la Recherche Scientifique (CNRS); Conseil
R\'egional Ile-de-France; D\'epartement Physique Nucl\'eaire et Corpusculaire
(PNC-IN2P3/CNRS); D\'epartement Sciences de l'Univers (SDU-INSU/CNRS);
Institut Lagrange de Paris (ILP) Grant No.~LABEX ANR-10-LABX-63 within
the Investissements d'Avenir Programme Grant No.~ANR-11-IDEX-0004-02;
Germany -- Bundesministerium f\"ur Bildung und Forschung (BMBF); Deutsche
Forschungsgemeinschaft (DFG); Finanzministerium Baden-W\"urttemberg;
Helmholtz Alliance for Astroparticle Physics (HAP);
Helmholtz-Gemeinschaft Deutscher Forschungszentren (HGF); Ministerium
f\"ur Innovation, Wissenschaft und Forschung des Landes
Nordrhein-Westfalen; Ministerium f\"ur Wissenschaft, Forschung und Kunst
des Landes Baden-W\"urttemberg; Italy -- Istituto Nazionale di Fisica
Nucleare (INFN); Istituto Nazionale di Astrofisica (INAF); Ministero
dell'Istruzione, dell'Universit\'a e della Ricerca (MIUR); CETEMPS Center
of Excellence; Ministero degli Affari Esteri (MAE); M\'exico -- Consejo
Nacional de Ciencia y Tecnolog\'\i{}a (CONACYT) No.~167733; Universidad
Nacional Aut\'onoma de M\'exico (UNAM); PAPIIT DGAPA-UNAM; The Netherlands
-- Ministry of Education, Culture and Science; Netherlands Organisation
for Scientific Research (NWO); Dutch national e-infrastructure with the
support of SURF Cooperative; Poland -Ministry of Science and Higher
Education, grant No.~DIR/WK/2018/11; National Science Centre, Grants
No.~2013/08/M/ST9/00322, No.~2016/23/B/ST9/01635 and No.~HARMONIA
5--2013/10/M/ST9/00062, UMO-2016/22/M/ST9/00198; Portugal -- Portuguese
national funds and FEDER funds within Programa Operacional Factores de
Competitividade through Funda\c{c}\~ao para a Ci\^encia e a Tecnologia
(COMPETE); Romania -- Romanian Ministry of Education and Research, the
Program Nucleu within MCI (PN19150201/16N/2019 and PN19060102) and
project PN-III-P1-1.2-PCCDI-2017-0839/19PCCDI/2018 within PNCDI III;
Slovenia -- Slovenian Research Agency, grants P1-0031, P1-0385, I0-0033,
N1-0111; Spain -- Ministerio de Econom\'\i{}a, Industria y Competitividad
(FPA2017-85114-P and FPA2017-85197-P), Xunta de Galicia (ED431C
2017/07), Junta de Andaluc\'\i{}a (SOMM17/6104/UGR), Feder Funds, RENATA Red
Nacional Tem\'atica de Astropart\'\i{}culas (FPA2015-68783-REDT) and Mar\'\i{}a de
Maeztu Unit of Excellence (MDM-2016-0692); USA -- Department of Energy,
Contracts No.~DE-AC02-07CH11359, No.~DE-FR02-04ER41300,
No.~DE-FG02-99ER41107 and No.~DE-SC0011689; National Science Foundation,
Grant No.~0450696; The Grainger Foundation; Marie Curie-IRSES/EPLANET;
European Particle Physics Latin American Network; and UNESCO.
\end{sloppypar}

%%%%%%%%%%%%%%%%%%%%%%%%%%%%%%%%%%%%%%%%%%
%%                                              BIBLIOGRAPHY                                                 %%
%%%%%%%%%%%%%%%%%%%%%%%%%%%%%%%%%%%%%%%%%%

\end{document}